\documentclass[3p,times,procedia]{elsarticle}
\usepackage{nupha_ecrc}

\volume{00}

\firstpage{1}

\journalname{Nuclear Physics A}

\runauth{}

\jid{nupha}

\jnltitlelogo{Nuclear Physics A}

\usepackage{amssymb}
 \usepackage{amsthm}

\newcommand{\trento}{T\raisebox{-.5ex}{R}ENTo}

\usepackage[figuresright]{rotating}
\usepackage{subcaption} 
\graphicspath{{fig/}}

\begin{document}

\begin{frontmatter}



\dochead{}

\title{A data-driven analysis of the heavy quark transport coefficient}

\author[rvt]{Yingru Xu}
\author[rvt,sub]{Marlene Nahrgang}
\author[rvt]{Jonah E. Bernhard}
\author[focal]{Shanshan Cao}
\author[rvt]{Steffen A. Bass}

\address[rvt]{Department of Physics, Duke University, Durham, NC 27708, USA}
\address[focal]{Department of Physics and Astronomy, Wayne State University, Detroit, MI, 48201}
\address [sub]{SUBATECH, UMR 6457, Universit´e de Nantes, Ecole des Mines de Nantes,
IN2P3/CNRS, 4 rue Alfred Kastler, 44307 Nantes cedex 3, France}

\begin{abstract}
Using a Bayesian model-to-data analysis, we estimate the temperature dependence of the heavy quark diffusion coefficients by calibrating to the experimental data of $D$-meson $R_{\mathrm{AA}}$ and $v_2$ in AuAu collisions ($\sqrt{s_{NN}}=200$ GeV) and PbPb collisions ($\sqrt{s_{NN}}=2.76$ TeV)~\cite{Xie:2016iwq}. The spatial diffusion coefficient $D_s2\pi T$ is found to be mostly constraint around $(1.3-1.5) T_c$ and is compatible with lattice QCD calculations. We demonstrate the capability of our improved Langevin model to simultaneously describe the $R_{\mathrm{AA}}$ and $v_2$ at both RHIC and the LHC energies, as well as the feasibility to apply a Bayesian analysis to quantitatively study the heavy flavor transport in heavy-ion collisions.
\end{abstract}


\end{frontmatter}

\section{Introduction}
\label{sec:intro}
Over the past few years, significant progress has been made to precisely describe the yields and flow of particles emitted from the QGP, and to quantitatively estimate its transport properties such as the shear viscosity to entropy density ratio $\eta/s$~\cite{Denicol:2015nhu,Bernhard:2016tnd,Novak:2013bqa}. High energetic jet and heavy quark energy loss mechanisms and their related transport coefficients ($\hat{q}, D_s$), however, are still not yet understood on a quantitative level. While the heavy quark transport coefficients are not directly measurable, they can be encapsulated in the parameters of theoretical models. By comparing the model calculations to the experimental data, we can estimate their values and therefore understand the interaction mechanism between heavy quarks and the medium.

In general, a comparison between the model calculations and the experimental data relies on optimizing multiple parameters -- some are related to the properties of the system, some are ad hoc parameters of different models. Until now, most heavy flavor studies vary the parameters manually until the agreement with the experimental data is achieved, which leads to very limited usefulness regarding comparison with experimental data. Moreover, the simultaneous description of various observables -- $D$-meson $R_{\mathrm{AA}}$ and $v_2$ -- is still an inevitable challenge for most models. A systematic and complete approach to optimizing the models would be to perform a random walk across the parameter space and calibrate to experimental data by applying a Bayesian analysis. In such an analysis the result is the posterior possibility distribution of each parameter, from which we can extract the optimal values of transport coefficients, the uncertainty and the correlation among different parameters.

The Bayesian statistical analysis has been applied with great success to the determination of multiple QGP properties, such as the precise estimate of the shear and bulk viscosities and  the constraint of the equation of state above the parton-hardon transition~\cite{Bernhard:2016tnd}. In this study, we extend the analysis to quantitatively study heavy flavor evolution in heavy-ion collisions.

\vspace{-0.3cm}
\section{Modeling heavy flavor evolution in heavy-ion collisions}
\label{s:model}
Our analysis utilizes the well-established framework developed by the Duke QCD group to describe the full space-time evolution of heavy quarks in heavy-ion collisions: the initial entropy density of the medium as well as the heavy quark position are generated by an effective parametric initial condition model \trento, which has been shown to mimic the scaling behavior of the EKRT and IP-Glasma models~\cite{Moreland:2014oya}. The initial transverse momentum distribution is provided by FONLL~\cite{FONLL}.

The propagation of the heavy quarks in the medium is described by an improved Langevin equation~\cite{Qin:2012fua}, which takes into account of both collisional and radiative energy loss:
\begin{equation}
\frac{d\vec{p}}{dt} = -\Gamma_D(p) \vec{p} + \vec{\xi} + \vec{f_g}
\end{equation}
The first two terms on the right hand side of the equation are the drag and thermal random forces inherited from the standard Langevin equation. They are responsible for the heavy quark collisional energy loss and related to the momentum diffusion coefficient $\hat{q}$ via  $\Gamma_D(p) = \hat{q}/(4TE)$ and $\left<\xi^i(t)\xi^j(t')\right> = \frac{1}{2} \hat{q} \delta^{ij} \delta(t - t')$. The third term $\vec{f}_g = -d\vec{p}_g/dt$ is the recoil force from the bremsstrahlung gluon emitted from the heavy quarks. It is added in order to take the radiative energy loss into consideration. We adopt the higher twist results for the gluon emission probability~\cite{Guo:2000nz}, which is proportional to the momentum diffusion coefficient $\hat{q}$. In literature~\cite{Qin:2012fua}, the spatial diffusion coefficient $D_s = 4T^2/\hat{q}$ is often used to characterize the interacting strength between the heavy quarks and the medium.

The evolution of the QGP medium is described by an event-by-event (2+1)-dimensional viscous hydrodynamical model VISH(2+1), which includes both shear and bulk viscous corrections~\cite{Song:2007ux}. It should be noted that all the parameters related to the bulk initialization and evolution have been calibrated to experimental data of charged particles~\cite{Bernhard:2016tnd}.

The hadronization of heavy quarks into heavy mesons is described by a hybrid fragmentation and recombination model. Currently we neglect any rescattering of the heavy mesons in the hadron gas state.

In order to estimate the heavy quark diffusion coefficient, we parametrize the spatial diffusion coefficient as linearly dependent on temperature and assume two model parameters $\vec{x}=(D_{\mathrm{min}}, D_{\mathrm{slope}})$:
\begin{equation}
D_s 2\pi T = D_{\mathrm{min}} + D_{\mathrm{slope}} (T - T_c)
\end{equation}
By varying the model parameters $\vec{x}$, we are able to change the temperature dependence of $D_s2\pi T$. In this work we focus on charm quarks evolution, but the same framework would apply to bottom quarks as well.

\vspace{-0.3cm}
\section{Model-to-data comparison}
To calibrate the model to the data, i.e. to determine the optimal values of all the parameters, requires a random walk through the parameter space using the Markov chain Monte Carlo (MCMC) algorithm\cite{Foreman:2012MCMC}. However, evaluating the full Langevin model in a fine grid in parameter space is computationally highly expensive, therefore directly performing the MCMC random walk with the Langevin model is not feasible. In this situation Gaussian process emulators can be used as an alternate surrogate model to fast interpolate the output of the Langevin model at any arbitrary point in the parameter space. In practice, a small number of parameter sets $((\vec{x}_1, \vec{x}_2,...,\vec{x}_n), i=1,...,80)$ are uniformly sampled via a Latin hypercube algorithm across the parameter space~\cite{Morris:2008lh} and evaluated in the Langevin model. The model outputs $\vec{y} = (R_{\mathrm{AA}}, v_2)$ are then calculated at each of the parameter set. Those prior results $((\vec{x}_i, \vec{y}_i), i=1,..,80)$ are used to train the Gaussian process emulators such that we can calibrate our parameters to experimental data through the MCMC random walk. The posterior possibility distribution $P(\vec{x}_*| X, Y, \vec{y}_{\mathrm{exp}})$ for the true parameter $\vec{x}_*$ is calculated according to Bayes' theorem~\cite{Novak:2013bqa}:
\begin{equation}
P(\vec{x}_*| X, Y, \vec{y}_{\mathrm{exp}}) \propto P(X, Y, \vec{y}_{\mathrm{exp}} | \vec{x}_*) P(\vec{x}_*)
\end{equation}
where $P(X, Y, \vec{y}_{\mathrm{exp}} | \vec{x}_*)$ is the likelihood of observing $(X, Y, \vec{y}_{\mathrm{exp}})$ with $\vec{x}_*$, and is given by~\cite{Pratt:2015zsa}:
\begin{equation}
P(X, Y, \vec{y}_{\mathrm{exp}} | \vec{x}_*) \propto \mathrm{exp}\left(-\frac{1}{2}(\vec{y}_* - \vec{y}_{\mathrm{exp}})^T \Sigma^{-1} (\vec{y}_* - \vec{y}_{\mathrm{exp}})\right)
\end{equation}

\begin{figure}
\centering
  {\includegraphics[width=0.9\textwidth, height=8cm]{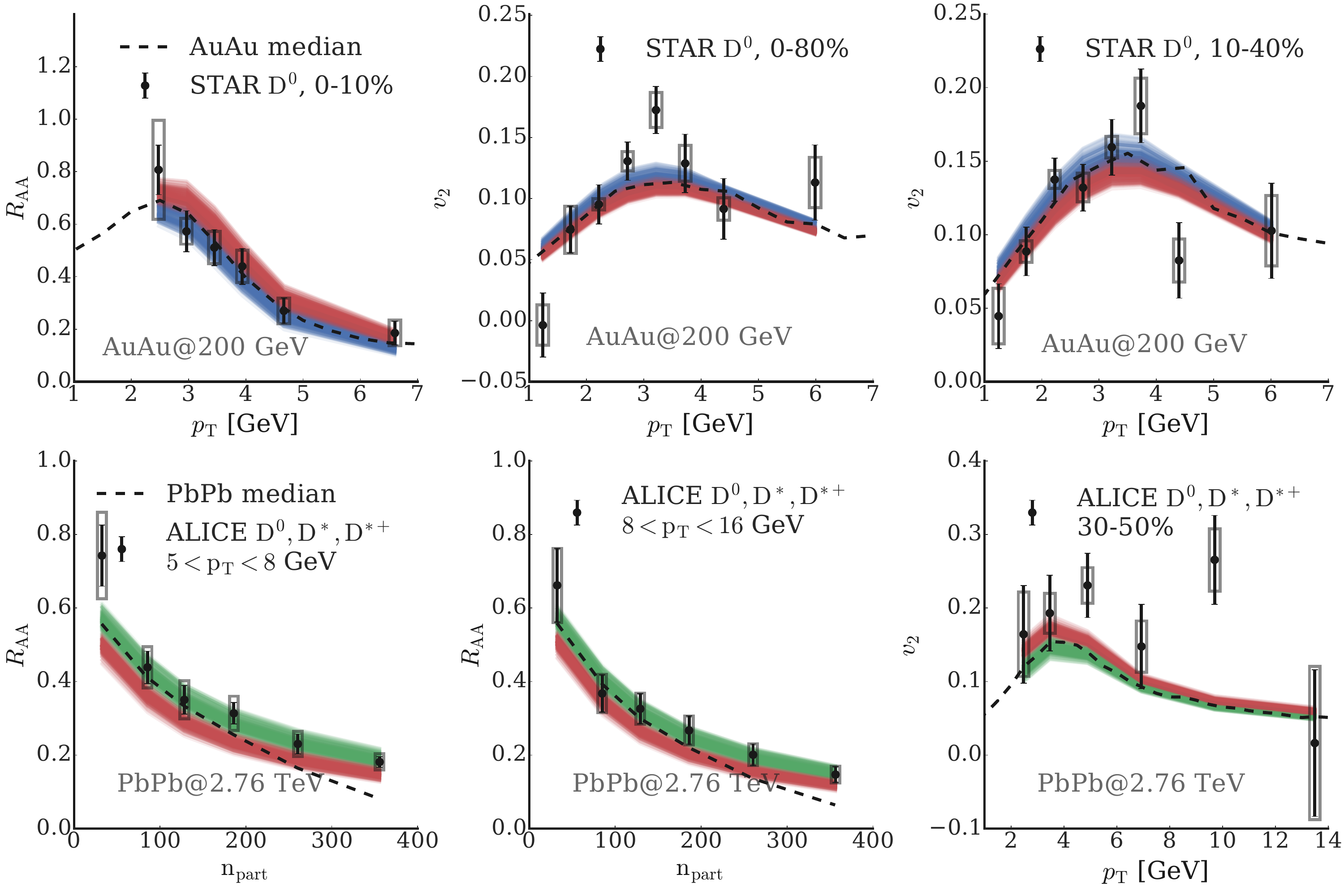}}
  \caption{(Color online) GP emulators prediction of 200 input samples randomly selected from the posterior distribution, and full Langevin calculation as taking the distribution median as parameters, compared with experimental data from STAR and ALICE~\cite{Xie:2016iwq}. }
  \label{f:Raa-v2}
\end{figure}

\vspace{-0.8cm}
\section{Posterior Results}
To verify our analysis, 200 input parameter values are randomly selected from the posterior distribution. In Fig.~\ref{f:Raa-v2} we compare our model outputs predicted from the Gaussian process emulator with the experimental data. Three individual analyses are performed in this work, and each is labeled with different color: the blue one calibrating our parameters to the experimental data in AuAu collisions, the green one calibrating to the experimental data in PbPb collisions, and the red one calibrating to both. The spread of the posterior outputs visualizes the remaining uncertainty in our analysis. In addition, the median value of each parameter $\vec{x}=(D_{\mathrm{min}},D_{\mathrm{slope}})$ is applied in the full Langevin simulation and the corresponding results are shown as the dashed lines. Overall the results demonstrate good agreement between the calibration and experimental values, as well as the validity of the Gaussian process emulator to predict the output from a physical model.

\begin{figure}
  \centering
  \begin{minipage}[b]{0.45\textwidth}
    \includegraphics[width=\textwidth]{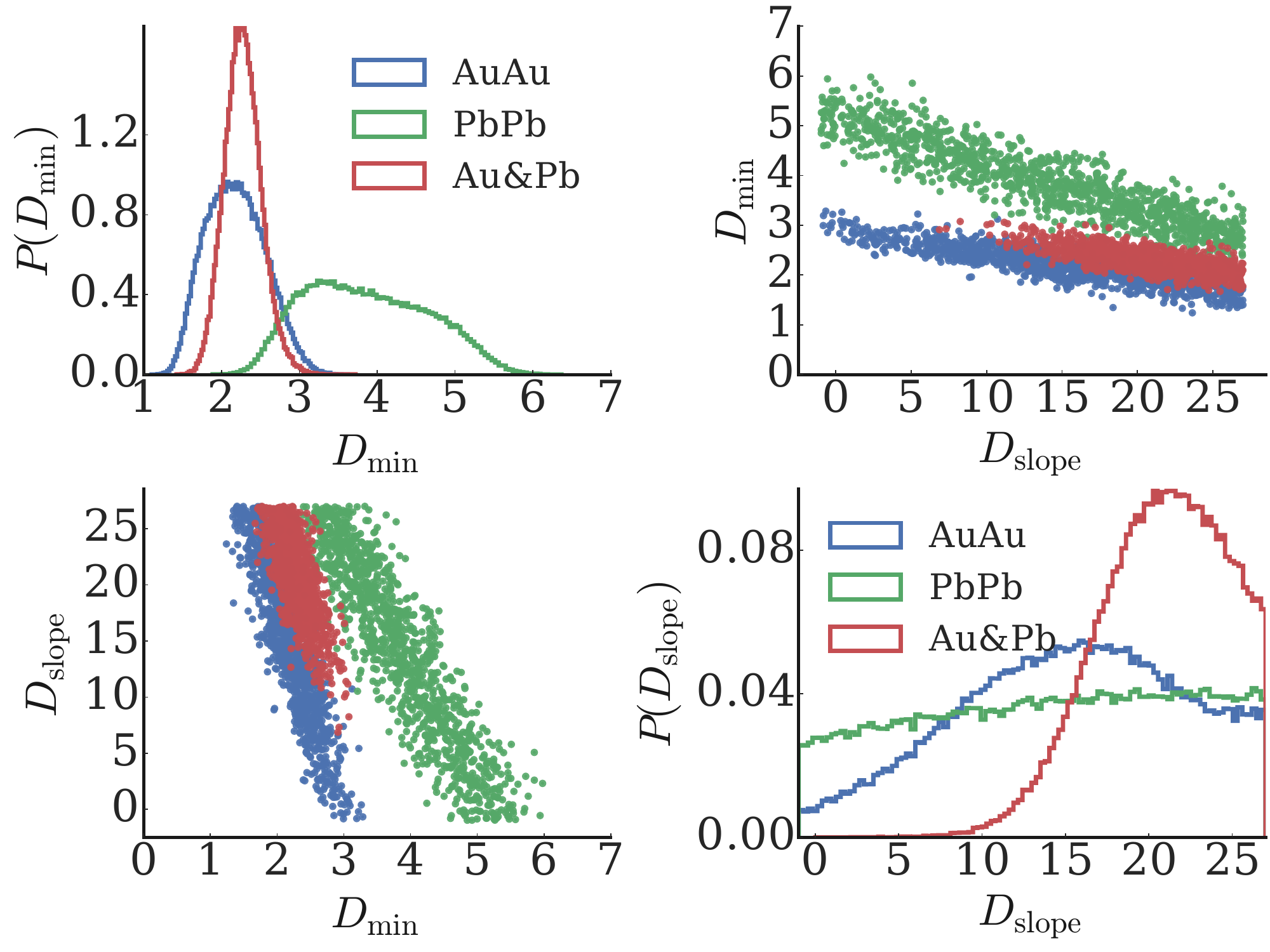}
    \caption{(Color online) Posterior distributions for the model parameters $(D_{\mathrm{min}}, D_{\mathrm{slope}})$, calibrated to STAR data of Au+Au collisions (blue), or calibrated to ALICE data of Pb+Pb collision (green), or calibrated to both (red).}
    \label{f:samples}
  \end{minipage}
\hfill
  \begin{minipage}[b]{0.45\textwidth}
    \includegraphics[width=\textwidth]{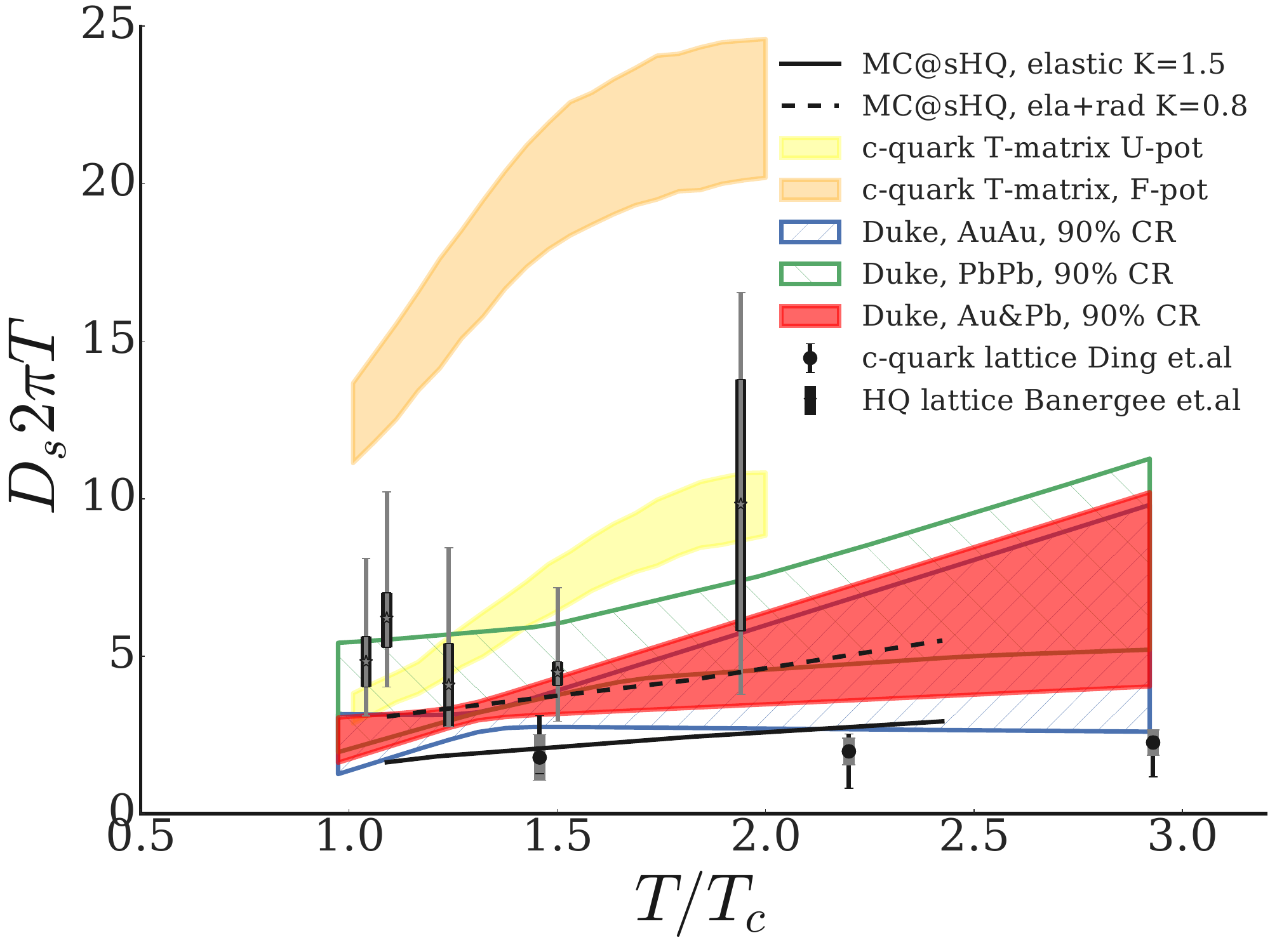}
    \caption{(Color onlie) Estimated temperature dependence of the spatial diffusion coefficient $D_s2\pi T$, compared with other models calculation. The hatched area are the 90\% credibility region (CR) from three calibrations.}
    \label{f:D2piT}
  \end{minipage}
\end{figure}

The main result of the analysis is the posterior possibility distribution of each parameter, which is shown in Fig.~\ref{f:samples}. The diagonal is the marginal distribution of the parameter $\vec{x}=(D_{\mathrm{min}}, D_{\mathrm{slope}})$ with the other one integrated out, and the off-diagonal shows the correlation between them. We find a narrow distribution for the parameter $D_{\mathrm{min}}$ for all the three analysis. From the off-diagonal histograms, we observe a non-trivial negative correlation between $D_{\mathrm{min}}$ and $D_{\mathrm{slope}}$. However, $D_{\mathrm{slope}}$ is relatively unconstrained, indicating a possible missing piece in our parametrization or too large experimental data uncertainties.

Figure ~\ref{f:D2piT} presents our estimate of the charm quark diffusion coefficient $D_s2\pi T$ as a function of temperature, compared to other model calculations~\cite{Ding:2012sp}. The posterior estimate for $D_s2\pi T$ is significantly constrained compared to the prior range. In addition, the uncertainty of the diffusion coefficient is smallest between $(1.3-1.5)T_c$ -- a temperature range where charm quarks spend most of the time. This feature of $D_s2\pi T$ is consistent with the Bayesian analysis on the temperature dependence of $\eta/s$, indicating that this is the temperature range where the medium spends most of the time. Though in general charm quarks tend to have a larger $D_s2\pi T$ in Pb+Pb collisions than in Au+Au collisions, the combined analysis still gives an estimate within the overlapped region of those two. As expected, the combined analysis is the most constrained. Our result is compatible with the lattice QCD calculations.

\vspace{-0.3cm}
\section{Conclusion}
In this study, we have performed a Bayesian model-to-data analysis to extract the temperature dependence of the charm quark diffusion coefficient in the QGP medium. The spatial diffusion coefficient $D_s2\pi T$ is found to be compatible to lattice QCD calculation, with the most constrained region between $1.3 T_c$ to $1.5 T_c$. This study has demonstrated the applicability of Bayesian analysis to the heavy flavor analysis in heavy-ion physics. Future work will improve the model description of the heavy quark evolution, such as applying Boltzmann transport, including hardonic scattering, and extend to more collision systems.

\textbf{Acknowledgements:} This work has been supported by the U.S Department of Energy under grant DE-FG02-05ER41367.  S.C is supported by the U.S DoE under grant DE-SC0013460. Computational resources were provided by the Open Science Grid (OSG), supported by DoE and NSF.





\vspace{-0.3cm}
\bibliographystyle{elsarticle-num}








\end{document}